\documentstyle[aps,prl,epsfig]{revtex}

\newcommand{\BE}{\begin{equation}}
\newcommand{\EE}{\end{equation}}

\def\bq{\begin{equation}}
\def\eq{\end{equation}}

\begin{document}
\title{Non-diffusive transport in plasma turbulence: a fractional diffusion
approach
\footnote{Research sponsored by Oak Ridge National Laboratory,
managed by UT-Battelle, LLC, for
the U.S. Department of Energy under contract DE-AC05-00OR22725.}}
\author{D. del-Castillo-Negrete
\thanks{e-mail: delcastillod@ornl.gov}\\
B. A. Carreras\\ V.~E. Lynch}
\address{Oak Ridge National Laboratory \\ Oak Ridge TN, 37831-6169}
\maketitle
 

\begin{abstract} 

Numerical evidence of non-diffusive transport  in three-dimensional, resistive
pressure-gradient-driven plasma turbulence is presented. It is shown that the
probability density function (pdf) of test particles' radial displacements is strongly non-Gaussian
and exhibits algebraic decaying tails.   To model these results we propose a
macroscopic transport model for the pdf  based on the use of fractional derivatives in space
and time, that incorporate in a unified way space-time
non-locality (non-Fickian transport), non-Gaussianity, and non-diffusive scaling. 
The fractional diffusion model reproduces the  shape, and space-time scaling of the non-Gaussian
pdf of turbulent transport calculations.  The model also reproduces the  observed super-diffusive
scaling.

\end{abstract} 
\vspace{0.75 cm}



Recent experimental and theoretical evidence indicates that transport in magnetically
confined fusion plasmas  deviates from the standard diffusion paradigm.  Typical examples include
the confinement  time scaling in low confinement mode plasmas
\cite{goldstone_1984,carreras_1997}, perturbative experiments
\cite{gentle_1995,callen_1997,cardozo_1995}, and the non-Gaussianity and long-range
correlations of fluctuations \cite{carreras_etal_1998}.   The standard diffusion paradigm breaks
down in these cases because it rests on restrictive assumptions including locality, Gaussianity,
lack of long-range correlations, and linearity.  In particular, according to Fick's law, the fluxes,
which contain the dynamical information of the transport process, are assumed to depend only on
local quantities, i.e. the gradients of the fields. Also, at a microscopic level, the diffusion
paradigm assumes the existence of an underlying un-correlated, Gaussian stochastic process, .i.e.  a
Brownian random walk.  

The need to develop models that go beyond these restrictive assumptions, is the main motivation of
this letter that has two connected goals. The first goal is to show numerical evidence of
non-diffusive transport in  pressure-gradient-driven plasma turbulence.  We do this by integrating
test particles in the ${\bf E}\times {\bf B}$ field obtained from a nonlinear, three-dimensional
 turbulence model. Test  particle  studies of this type have the advantage that incorporate in the
particle trajectories all the physics of the turbulence model.  However, this  ``microscopic"
approach has the limitation of being time consuming, and potentially redundant in the sense that it
tracks individual, particle orbit information  that from a statistical point view  might be
irrelevant.    This issue takes us to  the second goal of this letter which is to propose  and test a
macroscopic  model  describing the statistical properties of transport
in pressure-gradient-driven plasma turbulence.  The proposed model is based on the use of
fractional derivative operators which, as it will be explained below,  incorporate in  a natural,
unified way,  non-locality in space and time,  non-Gaussianity, and anomalous diffusion scaling.


The underlying  instability in pressure-gradient-driven plasma turbulence is the resistive
interchange mode, driven by the pressure gradient in regions where the magnetic field line
curvature is negative.  In this system, changes in the pressure gradient trigger instabilities at
rational surfaces that locally flatten the pressure profile and increase the gradient in nearby
surfaces. This in turn leads to successive instabilities and intermittent, avalanche-like  transport 
\cite{carreras_etal_1996}, which has been observed to cause anomalous diffusion 
\cite{carreras_etal_2001}. 
This instability is the analog of
the Raleigh-Taylor instability,  extensively studied in  fluids,
responsible for the gravity driven overturning of a low density fluid laying below a high density
fluid. In magnetically confined plasmas the role of gravity is played by
the curvature of the magnetic field lines which in a cylindrical geometry is always negative and
depends only on the radius.  

The turbulence model that we use, describes the coupled
evolution of the electrostatic potential $\Phi$ and pressure $p$  in a
cylindrical geometry \cite{carreras_etal_1996}
\bq
\label{turbulence}
\left(\frac{\partial }{\partial \tau}+ \tilde{\bf{V}}\cdot \nabla 
+\langle V_\theta\rangle \frac{1 }{r} \frac{\partial }{\partial \theta}
\right)\nabla_\perp^2\,
\tilde{\Phi}
 =
-\frac{1}{\eta m_i n_0 R_0}\nabla^2_\parallel \tilde \Phi+\frac{B_0}{m_i n_0}\frac{1}{r_c}
\frac{1}{r}\frac{\partial \tilde p}{\partial \theta}+ \mu \nabla^4_\perp \tilde \Phi
\eq
\bq
\left(\frac{\partial }{\partial \tau}+ \tilde{\bf{V}}\cdot \nabla 
+\langle V_\theta\rangle \frac{1 }{r} \frac{\partial }{\partial \theta}
\right)\tilde p=
\frac{\partial \left< p\right>}{\partial r} \frac{1}{r}\frac{\partial \tilde \Phi}{\partial \theta}
+\chi_\perp \nabla^2_\perp \tilde p+ \chi_\parallel \nabla_\parallel^2 \tilde p \, ,
\eq
where the tilde denotes fluctuating quantities (in time and space), and the angular bracket,
$\left< \, \right>$, denotes poloidal and toroidal angular (flux surface) average.  The magnetic
field 
$B_0$ is assumed to be on a cylinder with axis along the $z$-axis.
The equilibrium density is $n_0$, the ion mass is $m_i$, the averaged radius of curvature of the
magnetic field lines is $r_c$, and the resistivity is $\eta$.  
The subindex ``$\perp$" denotes the 
direction perpendicular to the cylinder's axis, and the subindex ``$\parallel$"  denotes the $z$
direction.  In both Eqs. (1) and (2) there are dissipative 
terms with characteristic coefficients $\mu$ (the collisional viscosity) and $\chi_\perp$ (the
collisional cross-field transport).  A parallel dissipation term proportional to
$\chi_\parallel$, is also included in the pressure equation.  This term can be interpreted as a
parallel thermal diffusivity. The evolution equation  of the flux surface averaged pressure is
\bq
\frac{\partial \left<p\right>}{\partial \tau}+ \frac{1}{r} \frac{\partial }{\partial r}r \left< \tilde
V_r
\tilde p\right>= S_0 + D \frac{1}{r}\frac{\partial}{\partial r}\left ( r \frac{\partial \left<
p \right>}{\partial r}\right) 
\eq
It contains a source term,  $S_0$, which is only a function of $r$.  This source of particles 
and heat is due, for instance, to neutral beam heating and fueling.  In this case, 
$S_0$ is essentially determined by the beam deposition profile.  In the present calculations, we
assume a parabolic  profile, $S_0=\bar{S}_0\left[1-(r/a)^2\right]$. 
The model parameters used here are $\mu = 0.2 \, a^2/\tau_R$ and $\chi_\perp = 0.025\,
a^2/\tau_R$, where
$\tau_R \equiv a^2\mu_0/\eta$ is the resistive time and $a$ the minor radius.  The rest of the
parameters in the model can be reduced to two dimensionless quantities the Lundquist number,
which is taken to be
$S = 10^5$, and  $\beta_0/2 \varepsilon^2=0.018$, where $\beta_0$ is
the value of $\beta$ at the magnetic axis, and $\varepsilon = a/R_0$.
The numerical calculations were carried out using the KITE code \cite{garcia_etal_1986}
with $363$ Fourier components to represent the poloidal and toroidal angle dependence for
each fluctuating component, and a  radial grid resolution of $\Delta r = 7.50\times10^{-4} a$.  

Having computed the electrostatic potential $\tilde \Phi$,  we study transport by following test
particle orbits determined from the solutions of  the ${\bf E}\times {\bf B}$ equation of motion
\bq
\label{test}
\frac{d {\bf r}}{d \tau}=
\frac{1}{B^2_0} \,\nabla \tilde
\Phi \times {\bf B}_0  \, .
\eq
Since the magnetic
field is fixed,  the turbulence-induce transport is only due to the fluctuating electrostatic
potential.   This electrostatic approximation, which is quite reasonable for low $\beta$ values,  is
needed  in order to carry out the numerical calculations in the time range required for reliable
transport studies. 
As  an initial condition we used $25,000$ tracer particles with random initial positions in $\theta$
and  $z$, and  radial position  $r = 0.5\, a$.  
Finite size effects did not seem to be relevant because during the evolution
there were very few particles moving out of the system. 
In the numerical integration of Eq.~(\ref{test}),  it is observed that  tracer particles either
get trapped in eddies for long times, or jump over several sets of eddies in a single flight,
giving rise to anomalous diffusion
\cite{carreras_etal_2001}.  

Due to the intrinsic stochasticity of  test particle orbits,  one has to resort to a
statistical approach to study transport in this system.  Our main object of study here is the
probability density function (pdf) of  radial displacements of the particles,
$P(x,t)$, where $x=(r-a/2)/a$ and $t=\tau/\tau_R$. By definition, at $t=0$, $P(x,t)=\delta x$. As
$t$ evolves, the pdf broadens and develop tails.  The triangles in Figure~1 show $P(x,t)$ at $t=0.64$
obtained from the histogram of particle displacements. 
The log-normal scale of the plot makes clear the strong non-Gaussianity of the density function (in
this scale a Gaussian is a parabola). The insert  in Fig.~1 shows that the tails exhibit algebraic
decay with exponent equal to $~1.75 \pm 0.03$.   
The numerical results show that for times above $t=0.1$, the moments of the test
particles displacements exhibits super-diffusive scaling, $\langle x^n\rangle \sim t^{n
\nu}$, with
$\nu = 0.66
\pm 0.02$. 


In what follows we show that these numerical results can be quantitatively
described with a transport model using fractional derivative operators in space and time.  The
generic form of the proposed  model is
\bq
\label{frac_diff_model}
 _{c}D_t^\beta P = \chi\,  \left[ w^-\,  _{a}D_x^\alpha + w^+\,  _{x}D_b^\alpha\right]\, P
+ \Lambda \, ,
\eq
where $\Lambda$ is a source, 
\bq
\label{eq_3}
_{a}D_x^\alpha P =\frac{1}{\Gamma(m-\alpha)} \, \partial^m_x\, \int_{a}^x\,  
\frac{P(y,t)}{\left(x-y\right)^{\alpha+1-m}}  \,dy\, , 
\eq
\bq
 _{x}D_b^\alpha P =\frac{(-1)^m}{\Gamma(m-\alpha)} \, \partial^m_x\, \int_{x}^b\,  
\frac{P(y,t)}{\left(y-x\right)^{\alpha+1-m}}  \,dy\, ,
\eq
are the left and right Riemann-Liouville fractional derivatives 
respectively,  $w^{\pm}$ are weighting factors,  and $m-1 \leq \alpha <m$
with $m$ a positive  integer.  The operator on the left hand side of Eq.~(\ref{frac_diff_model}) is
the Caputo fractional derivative in time of order $0< \beta<1$,
\bq
 _{c}D_t^\beta P=\frac{1}{\Gamma(1-\beta)} \, \int_{c}^t\,  
\frac{\partial_\tau P(x,\tau)}{\left(t-\tau\right)^{\beta}}  \,d\tau\, .
\eq
Despite the apparent complexity of their definition, fractional derivatives are natural
generalizations of regular derivatives. In particular, as expected, for $\alpha$ and $\beta$
integers,  these operators reduce  to regular derivatives, and  results of regular calculus
extend  directly to the fractional domain making the analytical study of fractional equations a
tractable problem. Further information about the definition and basic mathematical properties of
these operators can be found in Ref.~\cite{fractional_text}.

Fractional derivatives are integro-differential
operators that incorporate non-locality in space and time in a natural way. 
In particular,  the  right hand side of Eq.~(\ref{frac_diff_model}) evaluated at a  fixed position  $x$
takes into account non-local, spatial contributions to the flux from  all  points 
located to the left (through $_aD^\alpha_x$),  and all  points located to the right (through
$_xD^\alpha_b$) of $x$ \cite{paradisi_2001}.  The constants $w^\pm$  control the
degree of left-right asymmetry in the transport processes. This  is
relevant to fusion plasmas where asymmetric fluxes are usually observed.  The non-locality in time
is incorporated in the fractional derivative operator on the left hand side of
Eq.~(\ref{frac_diff_model}). Here, only the left derivative is used because, by causality, transport
can only depend on the past history of the system. 
In addition to the space-time non-locality, the fractional
diffusion model exhibits non-diffusive scaling of moments. 
In an infinite domain, the algebraic decaying tails of non-Gaussian distributions lead to divergent
moments. However, in physical applications (e.g. Ref.~\cite{metzler_klafter_2000}) a finite-$x$
cutoff leads to the finite size scaling $\langle  x^n \rangle \sim t^{n \nu}$, where 
$\nu=\beta/\alpha$.  Depending on the
value of $\alpha$ and $\beta$, transport can be super-diffusive ($2 \nu>1$), sub-diffusive
($2 \nu <1$), or diffusive ($2 \nu=1$).

The physics behind the model in Eq.~(\ref{frac_diff_model})  can be further understood from the
close connection  between transport models and the theory of random walks.  The standard diffusion
model is a macroscopic description of the Brownian random walk which  assumes that at fixed time
intervals
$t= T, \, 2 T,\, \ldots n T \ldots$ particles at a microscopic level  experience an
un-correlated  random displacement, or jump, $\ell_n$, with probability ${\cal P}_x$, where ${\cal
P}_x$ is assumed to have a finite second moment.   In a similar way, fractional diffusion models can
be viewed as macroscopic descriptions of  generalized  Brownian random walk models  known as  
the Continuous Time Random Walk (CTRW)  models \cite{metzler_klafter_2000}.  In addition to the
jump probability density ${\cal P}_x$, the CTRW model introduces a waiting time probability
function ${\cal P}_t$. That is,  the time between jumps, rather than being fixed as in a Brownian
walk, it is drawn from  a probability  function
${\cal P}_t$.  The different types of CTRW processes, and the resulting macroscopic transport
models,  can be classified based on the characteristic waiting time, $T$, and the characteristic
mean-square jump, $\sigma^2$, being finite or divergent \cite{metzler_klafter_2000}. 
Based on this, the
model (\ref{frac_diff_model}) involving fractional derivatives in space and time can be understood
as a general macroscopic description of an underlying microscopic stochastic process in which
particles exhibit  both, jumps without a characteristic spatial scale, and  waiting times without a
characteristic time scale.  The space non-locality is a direct consequence of the existence of
anomalously large jumps (known also as Levy flights) that connect  distant regions in space, and the
time non-locality is due to  the history-dependence introduced in the dynamics by the presence of
anomalously large waiting times.  

The fractional diffusion model in Eq.~(\ref{frac_diff_model}) is fairly general, and  depending on
the values of $\alpha$, $\beta$, and $w^{\pm}$, different transport processes can be modeled,
including  sub-diffusive transport, super-diffusive transport, and asymmetric transport.  In what
follows we show that for the symmetric, super-diffusive transport observed in
pressure-gradient-driven turbulence: $w^+=w^-=-0.5/ \cos(\pi \alpha/2)$,
$\alpha=3/4$,  $\beta=1/2$, and $\Lambda=0$.  To  understand this, consider the initial value
problem of Eq.~(\ref{frac_diff_model}) in an infinite domain, $x\in (-\infty,
\infty)$. Setting  $a=-\infty$ and $b=\infty$ in the  left and right fractional derivative operators,
and introducing the Fourier and Laplace transforms
\bq
\tilde{P}(k,t)=\int_{-\infty}^{\infty} P(x,t) e^{i k x} \, d x\, , \qquad 
\hat{P}(x,s)=\int_{0}^{\infty} P(x,t) e^{-s t} \, d t\, \, ,
\eq
Eq.~(\ref{frac_diff_model}) becomes
\bq
\label{transformed}
\left( s^\beta +  \chi \, |k|^\alpha \right)\tilde{\hat P}(k,s)= s^{\beta-1}
\tilde{P}(k,0)\, ,
\eq
where $\tilde{P}(k,0)$ is the Fourier transform of the initial condition,
and we have used the fact that $_{-\infty}D^\alpha_x\, e^{ikx}= (ik)^\alpha\, e^{ikx}$,
and $_{x}D^\alpha_{\infty}\, e^{ikx}= (-ik)^\alpha\, e^{ikx}$. The test particle transport studies
where done by releasing an ensemble of particles at a fixed radius. Based on this, we consider an
initial condition of the form  
$P(x,0)=\delta(x)$.
In this case, $P(x,t)$ becomes the Green's function or propagator which according to
Eq.~(\ref{transformed}) can be written as
\bq
\label{self_sim_sol}
P(x,t)= \frac{1}{2 \pi}\int_{-\infty}^\infty e^{-i k x}\, E_\beta(- \chi |k|^\alpha t^\beta) dk \, .
\eq
where 
\bq
 E_\beta(z)=\sum_{n=0}^{\infty} \frac{z^n}{\Gamma(\beta n+1)}\, ,
\eq
is the Mittag-Lefler function \cite{mainardi_etal_2001,fractional_text}.
As expected, for $\alpha=2$ and $\beta=1$, $P$ reduces to a Gaussian. For
$\beta=1$ , $1<\alpha\leq 2$, $P$ becomes a symmetric Levy stable distribution \cite{taqu}, and
for $0<\beta<1$ , $\alpha=2$ it reduces to the  solution of the sub-diffusion
fractional equation  \cite{metzler_klafter_2000}.  
Introducing the space-similarity variable, $\eta=t^{-\beta/\alpha} x$, the solution can be written
as
\bq
\label{sol_x}
P(x,t)= t^{-\beta/\alpha}\, K (\eta)\, ,  \qquad
K(\eta)= \frac{1}{\pi}\int_0^\infty \cos (\eta z)\,  E_\beta (-\chi z^\alpha) d z \, .
\eq
The solid line in Fig.~1 shows a  plot of  this solution  with $\beta=1/2$,
$\alpha=3/4$, and $\chi=0.09$.  The agreement with the test particles turbulence simulations
(triangles) is  good.  More precisely, 
using the asymptotic result 
$K(\eta) \sim \eta^{-(1+\alpha)}$ for large $\eta$ \cite{mainardi_etal_2001},
it follows that $P(x,t_0) \sim x^{-(1+\alpha)}$, for $x \gg t_0^{\beta/\alpha}$, 
which for $\alpha=3/4$ gives a decay exponent equal to $1.75$, a value in very good agreement with
the numerical result, $1.75 \pm 0.03$,  shown in the insert in Fig.~1.

The index $\beta$ determines the time-asymptotic  scaling properties of  $P$. 
To show this, we introduce the time-scaling variable $\zeta=t\,
|x|^{-\alpha/\beta}$, and write the solution as 
\bq
\label{sol_t}
P= |x|^{-1}\, \zeta^{-\beta/\alpha}\, K \left ( \zeta^{-\beta/\alpha}\right)\, . 
\eq
Using again the large $\eta$, and also the small $\eta$ asymptotic behavior of the function
 $K(\eta)$ it follows that $P(x_0,t) \sim t^\beta$, for $t \ll |x_0|^{\alpha/\beta}$, and 
$P(x_0,t) \sim t^{-\beta}$, for $t \gg |x_0|^{\alpha/\beta}$. 
This scaling is verified in Fig.~2 that shows
the evolution in time of $P$ at a fixed position $x_0$. The analytical solution according to
Eq.~(\ref{self_sim_sol}), shown with a solid line, exhibits  algebraic tails in the small $t$ and
large $t$ limits, and the expected peak at intermediate times. The circles and the triangles 
in the figure denote the results obtained from the test particles turbulence simulations. The
agreement is good, but  not as sharp as the one in Fig.~2 due to the numerical limitations in the
integration of the turbulence model for large times.  

As mentioned before,  $\alpha$ and $\beta$ determine the scaling properties of the 
moments of the test particles displacements. In particular,  $\langle x^n\rangle
\sim t^{n \nu}$, with 
$\nu= \beta/\alpha$. In the present case, $\alpha=3/4$ and $\beta=1/2$,  implies
$\nu=2/3$, a value in very good agreement with the one obtained from the test particles turbulence
simulation, $\nu = 0.66 \pm 0.02$.  The super-diffusive scaling
implies an anomalous confinement time scaling, $t_c \sim a^{\alpha/\beta}$.  For the case studied
here, $t_c \sim a^{3/2}$, a reasonable value in the range of the experimentally determined values 
which typically deviate from the standard-diffusion prediction $t \sim a^2$
\cite{carreras_1997}.

Summarizing, in this letter we have presented numerical
evidence that test particle transport in three-dimensional, resistive, pressure-gradient-driven
plasma turbulence exhibits non-diffusive transport. In particular, we have shown that the pdf of
particles is strongly non-Gaussian and exhibits algebraic tails with a decay exponent $1.75 \pm
0.03$.   Also, the moments of the test particles displacements exhibits supper-diffusive scaling
 with $\nu=0.66 \pm 0.02$.  We proposed a macroscopic
transport model for the pdf based on the use of fractional derivative
operators or order  $\alpha=3/4$ in space, and order $\beta=1/2$ in time. The model incorporates in
a natural, unified way, space-time nonlocality (non-Fickian transport), non-Gaussianity, and
anomalous diffusion scaling.  In  good agreement with the turbulent transport calculations, the pdf
in the fractional model decay with exponent $1+\alpha=1.75$, the pfd scale in time with exponent
$\beta=1/2$, and the moments scale with exponent $\nu=\beta/\alpha=2/3$ which implies a
confinement time scaling $t_c \sim a^{3/2}$. 
We have focused on symmetric fractional derivatives
(i.e. $w^+=w^-$). However,  the phenomenology of asymmetric operators is quite interesting, and
important in fusion plasmas.  For example, we have observed that asymmetric fractional derivative
operators give rise to  ballistic-like propagation of pulses. These  results indicate that fractional
diffusion models might be a useful tool to model rapid propagation phenomena in fusion devices.  
Another area where fractional operators looks promising is in the study of the role of 
non-diffusive transport in the L-H transition.   One way to approach this problem  is to incorporate
fractional diffusion operators into reaction-diffusion systems of the type used in
L-H transition studies (e.g. Ref.\cite{del_castillo_2002}).   A first step in this direction 
was  presented in Ref.~\cite{del_castillo_2003} where it was shown that fractional diffusion gives
rise to asymmetric, exponential acceleration of fronts.

\nopagebreak
%

%




%
%

%

\begin{figure}
\epsfig{figure=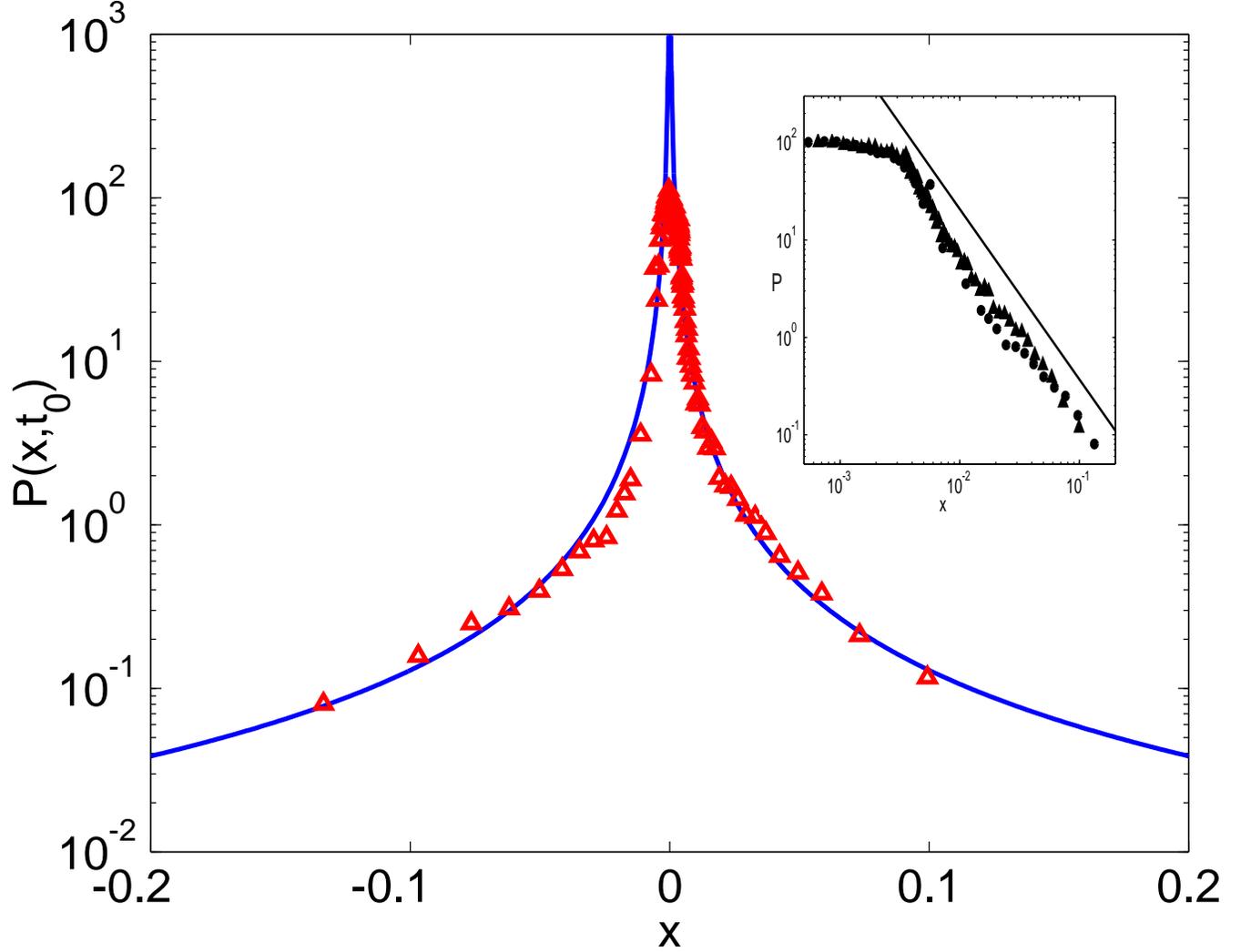,width=\columnwidth,angle=0}
\caption{Non-Gaussian probability density function of test particles in plasma
turbulence. The triangles denote the results from the histogram of radial displacements according
to the test particle, pressure-gradient-driven turbulence model 
Eqs.~(\ref{turbulence})-(\ref{test}). The solid line is the analytical solution  in Eq.~(\ref{sol_x})
of the symmetric ($w^+=w^-$) fractional diffusion transport model in Eq.~(\ref{frac_diff_model})
with $\alpha=3/4$,
$\beta=1/2$ and $\chi=0.09$.  The log-log insert shows the algebraic decay of the left (circles) and
right (triangles) tails. The straight line in the insert is a fit with the predicted decay exponent,
$1+\alpha=7/4$}
\label{fig_3}
\end{figure}


\begin{figure}
\epsfig{figure=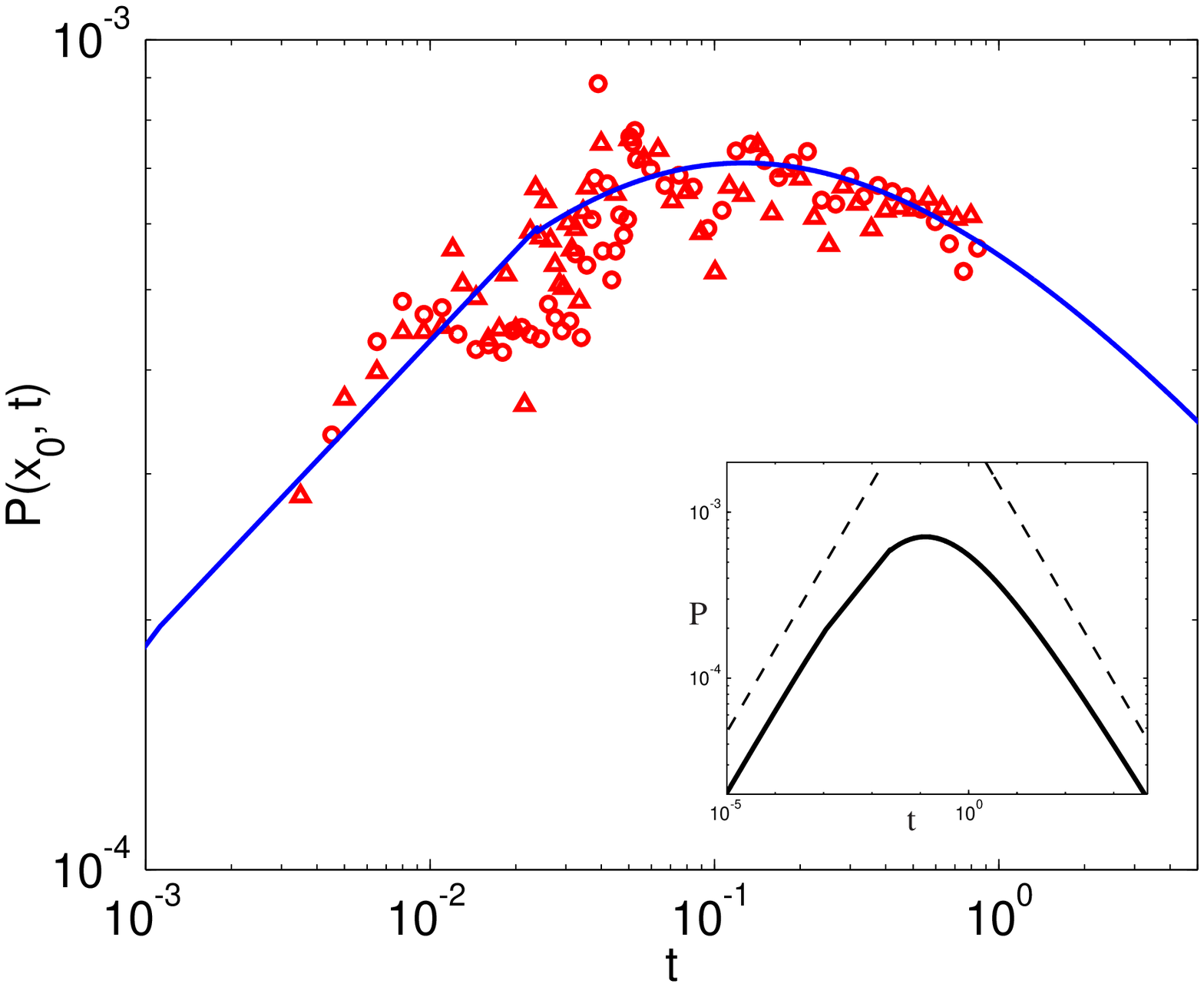,width=\columnwidth,angle=0}
\caption{Time evolution of the probability density function of test particles in
pressure-gradient-driven plasma turbulence.   The circles and the triangles denote the results from
the turbulence model  Eqs.~(\ref{turbulence})-(\ref{test}). The solid
line is the analytical solution   (\ref{sol_t}) of the symmetric ($w^+=w^-$) fractional diffusion
transport model in Eq.~(\ref{frac_diff_model}) with
$\alpha=3/4$, $\beta=1/2$ and $\chi=0.09$.  In agreement with the asymptotic result, the insert
shows that the pdf exhibits algebraic tails with exponent equal to
$\beta=1/2$. }
\label{fig_4}
\end{figure}



\begin{thebibliography}{99}
%

\bibitem{goldstone_1984}
R.~J. Goldstone. Plasma Phys. Controlled Fusion {\bf 26}, 87 (1984). 

\bibitem{carreras_1997}
B.~A. Carreras, IEEE Transactions of Plasma Science, {\bf 25} 1281 (1997). 

\bibitem{gentle_1995}
K. Gentle, G. Cima, H. Gasquet, G.~A. Hallock {\em et al.}, Phys. Plasmas {\bf 2}, 2292 (1995). 

\bibitem{callen_1997}
J.~D. Callen, Plasma Phys. Controlled Fusion, {\bf 39}, B173 (1997). 

\bibitem{cardozo_1995}
N. Lopez Cardozo, Plasma Phys. and Controlled Fusion {\bf 37}, 799 (1995). 

\bibitem{carreras_etal_1998}
B.~A. Carreras, B. v. Milligen, M.~A. Pedroza {\em et al.}, Phys. Rev. Lett. {\bf 80}, 4438 (1998). 

\bibitem{carreras_etal_1996}
B.~A. Carreras, D. Newman, V.~E. Lynch, {\em et. al}, Phys. Plasmas. 3, 2903 (1996).

\bibitem{carreras_etal_2001}
B.~A. Carreras,  V.~E. Lynch, {\em et al}, Phys. Plasmas {\bf 8}, 5096
(2001).

\bibitem{garcia_etal_1986}
L. Garcia, {\em et al.}, J. Comput. Phys. {\bf 65}, 253 (1986).

\bibitem{fractional_text} I. Podlubny, {\em Fractional Differential Equations} (Academic Press, San
Diego, 1999). 

\bibitem{paradisi_2001}
P. Paradisi, R. Cesari, F. Mainardi, and F. Tampieri,
Physica A, {\bf 293}, 130-142 (2001). 

\bibitem{metzler_klafter_2000}
R. Metzler, and J. Klafter, Phys. Rep., {\bf 339}, 1, (2000).


\bibitem{mainardi_etal_2001}
F. Mainardi, Y. Luchko, and G. Pagnini,
Fractional Calculus and Applied Analysis, {\bf 4}, 153-192 (2001). 

\bibitem{taqu}
G. Samorodnitsky, and M.~S. Taqqu, {\em Stable non-Gaussian random processes}
(Chapman \& Hall, New York, 1994). 


\bibitem{del_castillo_2002}
D. del-Castillo-Negrete, B.~A. Carreras, and V. Lynch, 
Phys.  Plasmas {\bf 9}, 118, (2002). 

\bibitem{del_castillo_2003}
D. del-Castillo-Negrete, B.~A. Carreras, and V. Lynch, Phys. Rev.
Lett. {\bf 91}, 018302-1, (2003). 

\end{thebibliography}
\end{document}